\documentstyle[12pt]{article}
\input epsf.tex
\epsfclipon
\newcommand{\nc}{\newcommand}
\nc{\beq}{\begin{equation}}
\nc{\eeq}{\end{equation}}
\nc{\beqa}{\begin{eqnarray}}
\nc{\eeqa}{\end{eqnarray}}
\nc{\lra}{\leftrightarrow}
\nc{\la}{\lambda}

\nc{\sss}{\scriptscriptstyle}
\nc{\lsim}{\mbox{\raisebox{-.6ex}{~$\stackrel{<}{\sim}$~}}}
\nc{\gsim}{\mbox{\raisebox{-.6ex}{~$\stackrel{>}{\sim}$~}}}

\begin{document}

\begin{flushright}
McGILL-03-01\\
CERN-TH/2003-004
\end{flushright}

\vskip.5cm
\begin{center}
{\huge{\bf Real-time D-brane condensation}}
\end{center}
\vskip0.5cm

\centerline{ James M.\ Cline$^{*,\dagger}$ and Hassan Firouzjahi$^\dagger$}
\bigskip
\centerline{$^*$ Theory Division, CERN, CH-1211, Geneva 23, Switzerland}
\centerline{$^\dagger$ Physics Department, McGill University,
Montr\'eal, Qu\'ebec, Canada H3A 2T8}

\begin{abstract}
\vskip 3pt
Unstable D-branes or brane-antibrane systems can decay to lower-dimensional branes.  In
the effective field theory description, the final state branes are defects in the tachyon
field which describes the initial instability.  We study the dynamical formation of
codimension one defects (kinks) using Sen's ansatz for the tachyon Lagrangian.   It is
shown that the slope of the kink diverges within a finite amount of time after the
tachyon starts to roll.  We discuss the relevance for reheating after brane-antibrane
inflation.
\end{abstract}

{\bf 1.\ Introduction.} In the last few years, significant progress has been made in
understanding the nature of transitions in unstable systems of D-branes \cite{Sen}.  Most
of the work has been motivated by formal considerations, but there is also an important
application, namely inflation from brane-antibrane collisions
\cite{brane-inflate,angles}.  This is an appealing application of genuine string
theoretic constructions to cosmology, but it may have problems with reheating
\cite{Kofman-Linde, stw, CFM}.  It seems likely that the energy density liberated from the
brane collisions is efficiently converted into closed string states (ultimately
gravitons) \cite{Strominger, Dvali-Vilenkin}, and not necessarily into visible
radiation.  To quantitatively address this potential problem, one should consider how the
evolving tachyon condensate couples to photons, which being an open string excitation
should reside on some stable brane remaining in the final state.

One way in which branes could be created in the collision is through the Kibble mechanism
\cite{sarangi-tye, majum-davis}; the tachyon field forms topological defects, which are known to be a
consistent description of branes whose dimension is lower than that of the original
branes \cite{Sen-descent}.  For example, a brane-antibrane system has a complex tachyon
field, leading to vortices which represent codimension-two branes.  On the other hand an
unstable brane has a  real tachyon field, which can form kinks in some direction,
representing codimension one branes.  For simplicity we are going to study the latter
situation in this letter.  It should be noted however that this process could also
originate from a brane-antibrane collision, in which one component of the complex tachyon
first undergoes condensation to form unstable codimension-one kinks, followed by the
second, orthogonal component; the intersection of these two kinks is the vortex.  Our
simplified situation describes the second step in this process.

The static properties of tachyon defects have been well-studied in the literature; the
fact that their tensions match the known ones of D-branes is part of the evidence that
they {\it are} D-branes \cite{KMM,kraus-larsen}.  In addition, excitations around these defects have been shown
to reproduce the excited states of strings in the case of $p$-adic strings, where such
calculations are tractable \cite{min-zwie}.  However, not much is known about the dynamics of defect
formation; precisely how do they form in space and time, starting from an unstable
configuration?  These details will be important for making a quantitative calculation
of reheating on the branes which form (one of which is presumed to contain the standard
model).  We may then hope to settle the question of whether such a brane universe will be
completely dominated by gravitational radiation at the end of inflation.

In the present work, we do not attempt to improve on the reheating computation,
contenting ourselves with determining the tachyon profile $T(t,x)$ as a function of time
and the extra dimension which is transverse to the kink.  We will show that Sen's version
of  the effective action for the tachyon leads to a somewhat surprising result: the slope of
the kink diverges in a finite time after the brane collision.  We show that this is
related to the observation that caustics can also form in this situation.  We conclude by
speculating that both this problem and that of the caustics is an artifact of ignoring
higher derivative corrections to the tachyon action.

{\bf 2.\ Action and equations of motion.}  
Sen has proposed a simple form for the tachyon action which captures the essential
qualitative features of exact computations from boundary string field theory (BSFT),
and which quantitatively agrees with the exact result in certain limits \cite{Garousi,
Sen-action}. It has the
form
\beq 
\label{action1}
  {\cal L} = -{\cal T} V(T)
  \sqrt{-\det(g_{\mu\nu} - \partial_\mu T\partial_\nu T )}
\eeq
where ${\cal T}$ is the tension of the original nonBPS brane, and 
the potential is $V(T) = e^{-|T|^2/a^2}$ for the superstring.  $T$ has
dimensions of length and $a$ is of order the string length scale.
Evaluating the determinant, and assuming $T$ varies in only one spatial
dimension $x$, which we shall refer to as the bulk, the action becomes
\beq 
\label{action2}
  {\cal L} = -{\cal T} V(T)
  \sqrt{1-\dot T^2 + T'^2 }
\eeq
Let us compare this with the exact result from BSFT \cite{KMM,kraus-larsen}:
\beq
\label{action3}
  {\cal L} = -{\cal T} e^{-T^2/a^2} F[-(\partial_\mu T)^2]
\eeq
where 
\beq
\label{Feq}
	F(x) = {4^x x \Gamma(x)^2\over 2\Gamma(2x)},
\eeq
Although this looks very different from (\ref{action2}), it has some essential
similarities. For example both $F(x)$ and $\sqrt{1+x}$ approach $\sqrt{x}$ in
the limit of large $x$, appropriate for describing the static kink solution,
which has the property that $T'\to\infty$. Moreover, both functions have first
derivatives which diverge as $x\to-1$, which puts a limit on how fast the
field can roll for homogeneous configurations: $\dot T\to \pm 1$.  (It turns out
that the zeroes of $F(x)$ and $\sqrt{1+x}$ do not have special significance,
so it is not important that they do not coincide.)  We will work with Sen's
form of the action since it captures the distinctive behavior of the exact one,
but is much simpler to work with.  We checked certain of the following
numerical results also using the BSFT action to be sure that no significant
differences arose.

The equation of motion which follows from (\ref{action3}) is
\beq
\label{eom}
\ddot T = 
{\frac {2a^{-2}T(1-\dot T^2 + T'^2)+(1-\dot T^2)T'' +2\dot T T' \dot T'}
{\left( 1+T'^{2} \right) }}
\eeq
In the homogeneous case, this simplifies to $\ddot T = 2a^{-2} T(1-\dot T^2)$,
which shows the limiting velocity $\dot T\to 1$.  The asymptotic form of the
solution is 
\beq
\label{homo_soln}
T\cong T_i + t - {\sqrt{2\pi} a\over 8}{\rm erf}({\sqrt{2}t\over a})
\eeq
at large times.  

{\bf 3.\ Dynamical solutions.}   For nonhomogeneous tachyon configurations in
which $T$ has reached large values $T\gg a$, analytic approximations to the
solution have been developed in reference \cite{FKS}.  However in the present
study we are interested in formation of kinks, near which $T$ remains zero.
Before looking for analytic approximations, let us consider numerical evolution of
the equation of motion (\ref{eom}).  We assume the extra dimension is
compactified on a circle of circumference $2L$, and approximate it by $N$
discretize points. We  put periodicd boundary conditions on $T$, and replace
spatial derivatives in eq.\ (\ref{eom}) by finite
differences. This results in a set of $N$  coupled ordinary differential
equations which can be solved numerically.

The initial condition is taken to be $T(0,x)= \epsilon\cos(\pi x/L)$ where
$\epsilon\ll a$.  This is an idealization of random initial conditions where
$T(0,x)$ happened to cross zero at two locations in the compact bulk. Fig.\ 1
shows the how the spatial profiles evolve in time.
\vskip -0.5cm
\centerline{\epsfxsize=4.5in\epsfbox{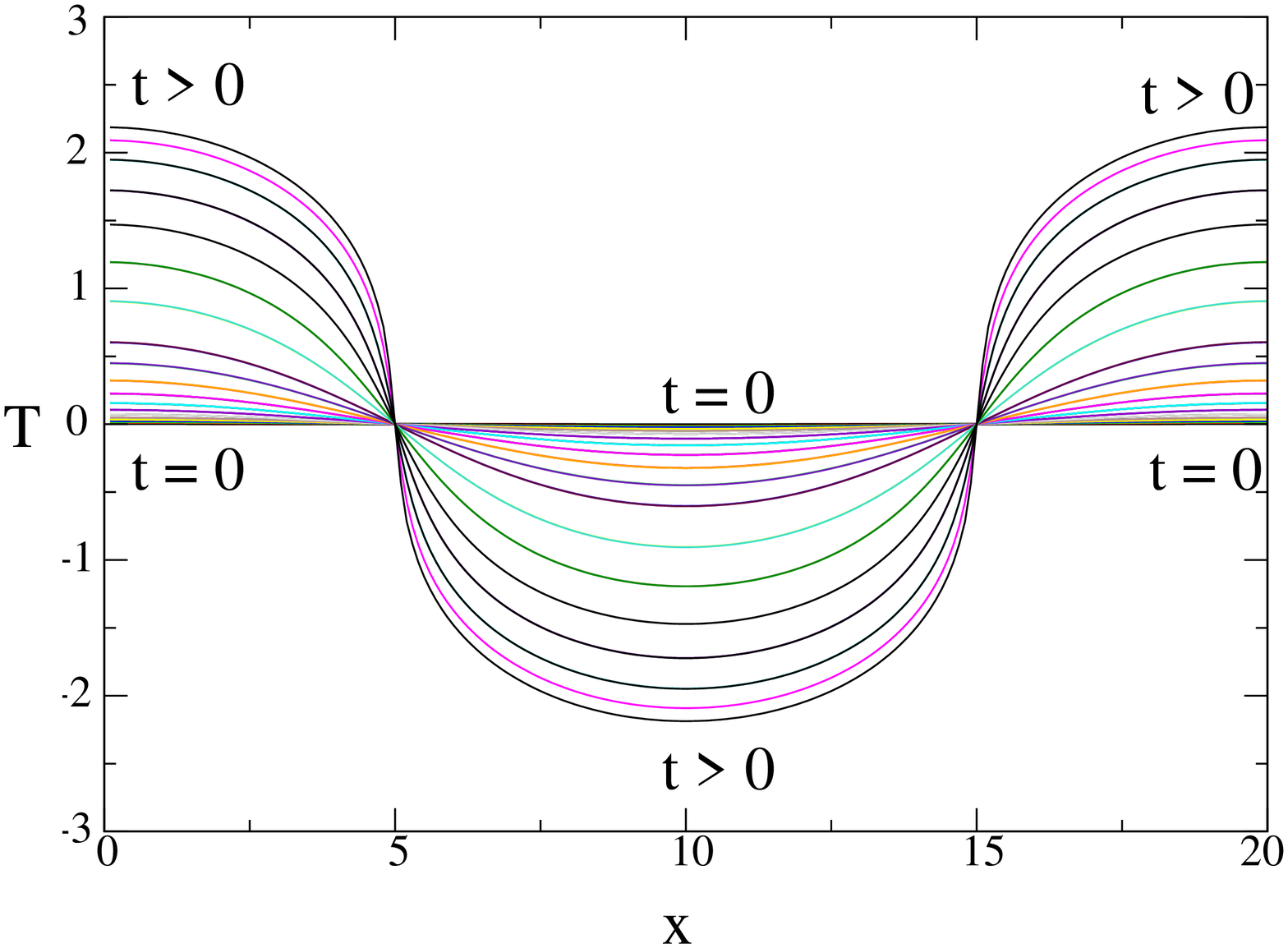}}
{\small
Figure 1. Sequence of spatial tachyon profiles $T(x)$ for a series of
increasing times, showing formation of a kink-antikink pair.
}

A few features of the above solutions are noteworthy.  At any position $x$
apart from the kink locations, $T(t,x)$ approaches the behavior
(\ref{homo_soln}), where $T_0 = T_0(x)$.  This is demonstrated in Fig.\ 2,
which plots $\dot T(x)$ for a sequence of increasing times.  It is clear
that $|\dot T|\to 1$ at any $x$, though the approach to the asymptotic value
takes longer for points closer to the kink locations.  Secondly, the slope
of the kink becomes quite steep on a short time scale.  In fact, the code
crashes at a certain finite time because the slope becomes singular.

\medskip
\centerline{\epsfxsize=4.5in\epsfbox{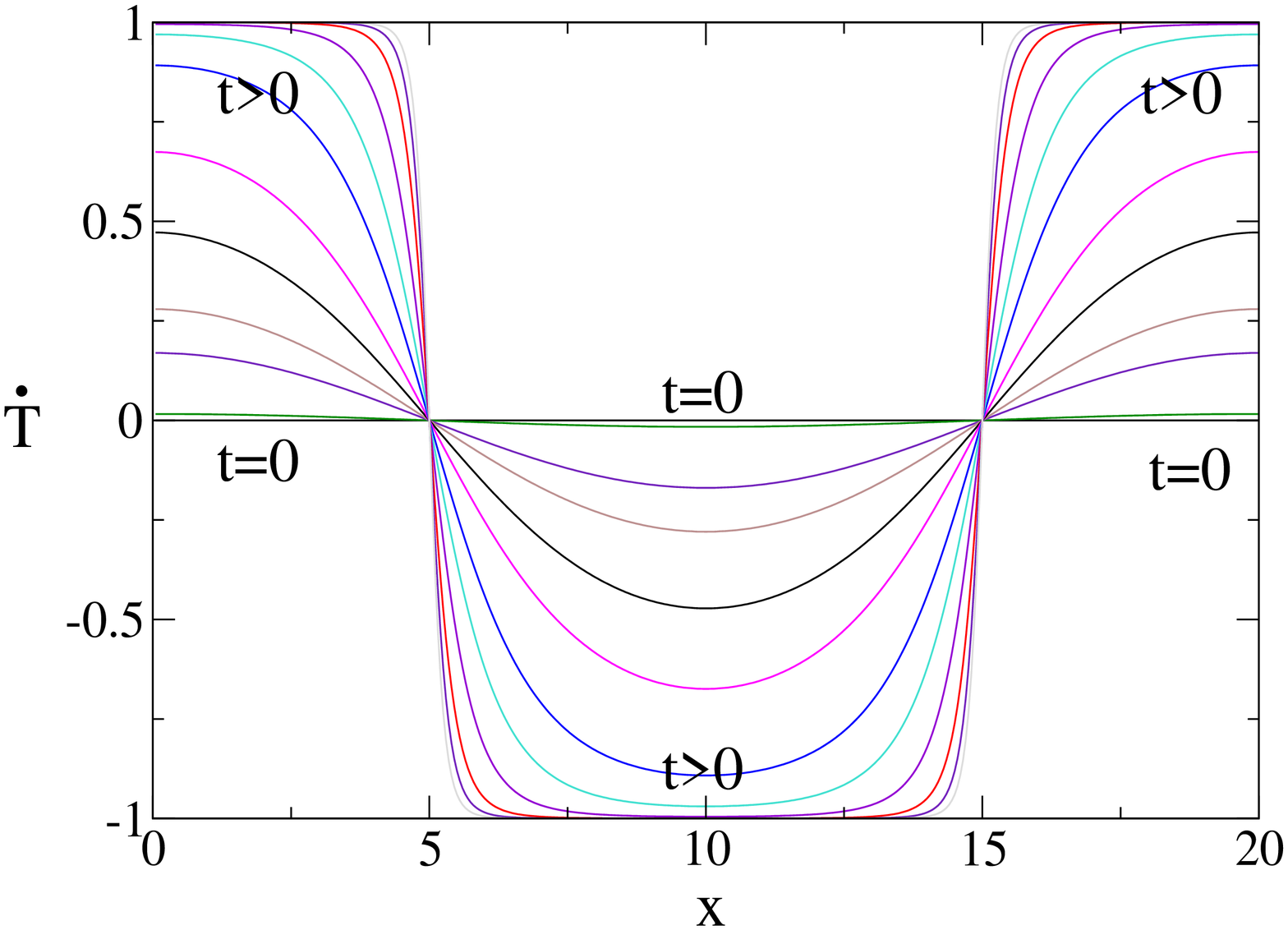}}
{\small
Figure 2. Sequence of spatial tachyon velocity profiles $\dot T(x)$ for a series of
increasing times, showing that $|\dot T|\to 1$ in the bulk.
}
\medskip

One might wonder if numerical error is the reason that the evolution cannot
be followed past a certain time.  However this seems not to be the case.
Using finer lattices and time steps did not help the problem.  Neither did
higher order differencing schemes for discretizing the spatial derivatives. It
is interesting to see in detail why the numerical evolution fails beyond the
critical time.  To show this, consider $T(t,x_i)$ at the lattice sites  $x_i$
which are in the vicinity of one of the kinks.  The behavior of the quantity
$1-\dot T^2 + T'^2$ which appears in the tachyon action is shown in Fig.\ 3. 
The curious feature is that at the lattice sites neighboring the position of
the kink (on either side), $1-\dot T^2 + T'^2$ starts plunging toward negative values after a
certain time.  Such values are unphysical since the square root of
 $1-\dot T^2 + T'^2$ appears in the action, and this is what causes the numerical
evolution to crash.  Again, one might suspect some sort of numerical
problem, but this behavior proved to be completely insensitive to all manner of
modifications to the program which were tried.  We will give some analytical
arguments for why this pathological end to the evolution is in fact inherent
in the equations of motion, rather than a fluke of the numerics.
\medskip
\centerline{\epsfxsize=4.5in\epsfbox{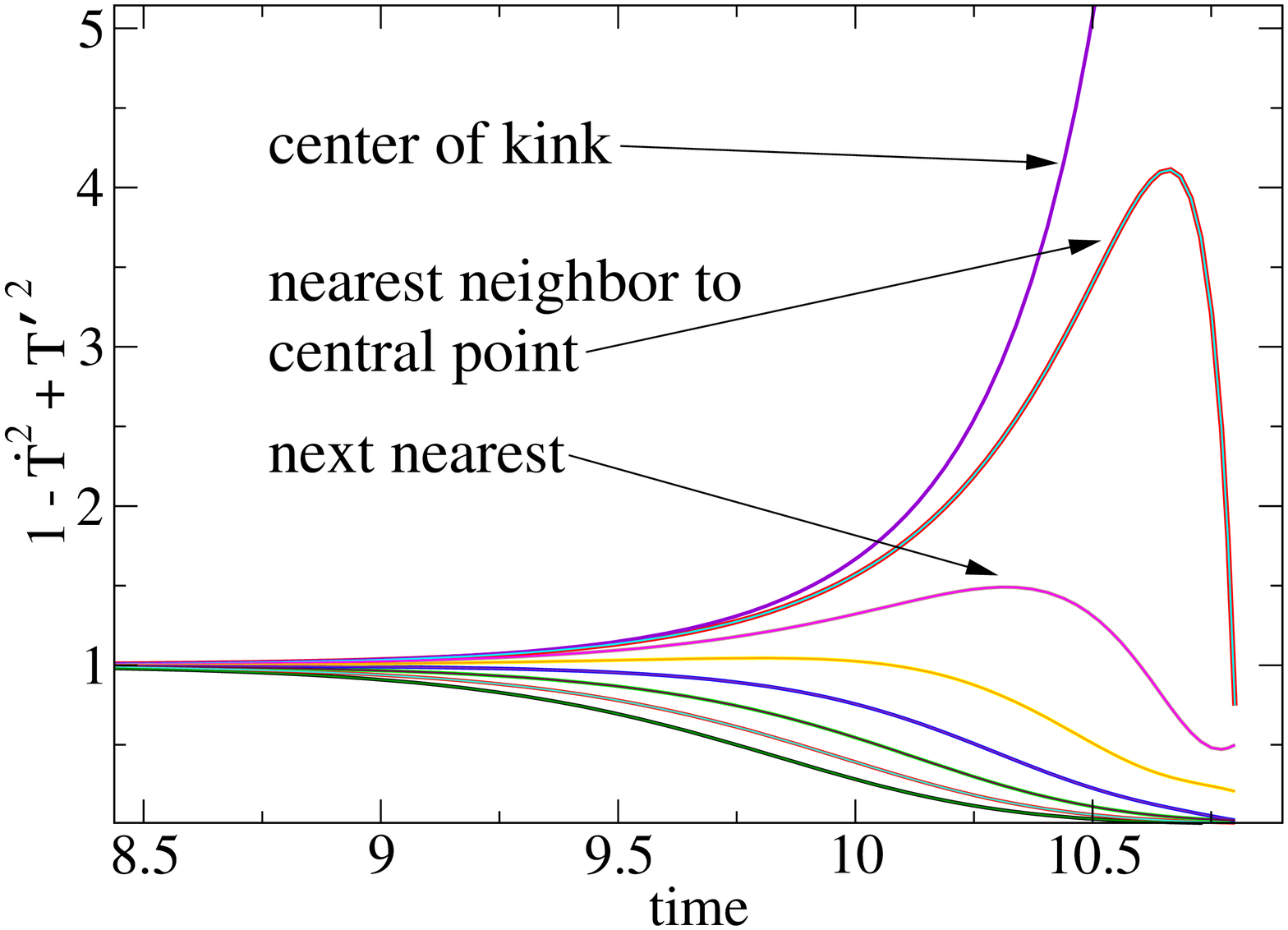}}
{\small
Figure 3. The quantity $1-\dot T^2 + T'^2$ as a function of time in units
of $a$, for
several lattice sites in the vicinity of a kink.  The rapid descent of
the curve for the nearest neighbor to the kink center signals a pathology
in the time evolution.}
\medskip

{\bf 4.\ Analytic approach.} To analytically study the dynamics of $T(t,x)$ close
to the kink, we can use the fact that it is odd in $x$ and write the ansatz
\beq
\label{Texp}
	T(t,x)\cong q(t) x + p(t) x^3 + r(t) x^5 + \cdots
\eeq
Substituting this into the equation of motion and ignoring the terms which
are subleading in $x$ gives
\beq
\label{qeq}
	\ddot q = {2\over a^2} q + {2 q \dot q^2 + 6p\over 1+q^2}
\eeq
This cannot be solved for $q$ since it depends on $p$, whose solution depends
on $r$, {\it etc.}, but let us suppose that $p$ is initially zero and can
thus be ignored at least for early times.
The solution has two regimes.  At early times, before $\dot q$ has become large,
the second term is negligible and therefore
\beq
\label{qexp}
	q(t) \cong q_+ e^{\sqrt{2}t/a} + q_- e^{-\sqrt{2}t/a}
\eeq
However $\dot q$ quickly grows, so for $t\gg a$ it will no longer be consistent
to ignore the second term in (\ref{qeq}).  One can show that the solution in
the regime where the second term dominates has the behavior
\beq
\label{qdiv}
	 q \sim {c\over t-t_0 }
\eeq
This behavior is borne out by the numerical solutions, as shown in Fig.\ 4,
which plots $\ln(T(t,x_i)/x_i)\cong q(t)$ versus $t$ for the lattice points
closest to the kink.  At early times the kink formation starts very slowly
(both terms in (\ref{qexp}) are important), while the
 intermediate straight-line behavior confirms
(\ref{qexp}) when the positive exponential dominates.  After a time
interval of approximately $2a$ from the start of the linear region,
the curves start to diverge from each other, showing that $T$ is no longer just
linear in $x$ over the range of $x_i$ given.  But the topmost curve is given
by the site $x_1$ which is the nearest neighbor to the kink, so this gives
the best approximation to the linear term in (\ref{Texp}), and we see that
it begins to rise faster than exponentially with time, in qualitative 
agreement with (\ref{qdiv}).

\medskip
\centerline{\epsfxsize=4.5in\epsfbox{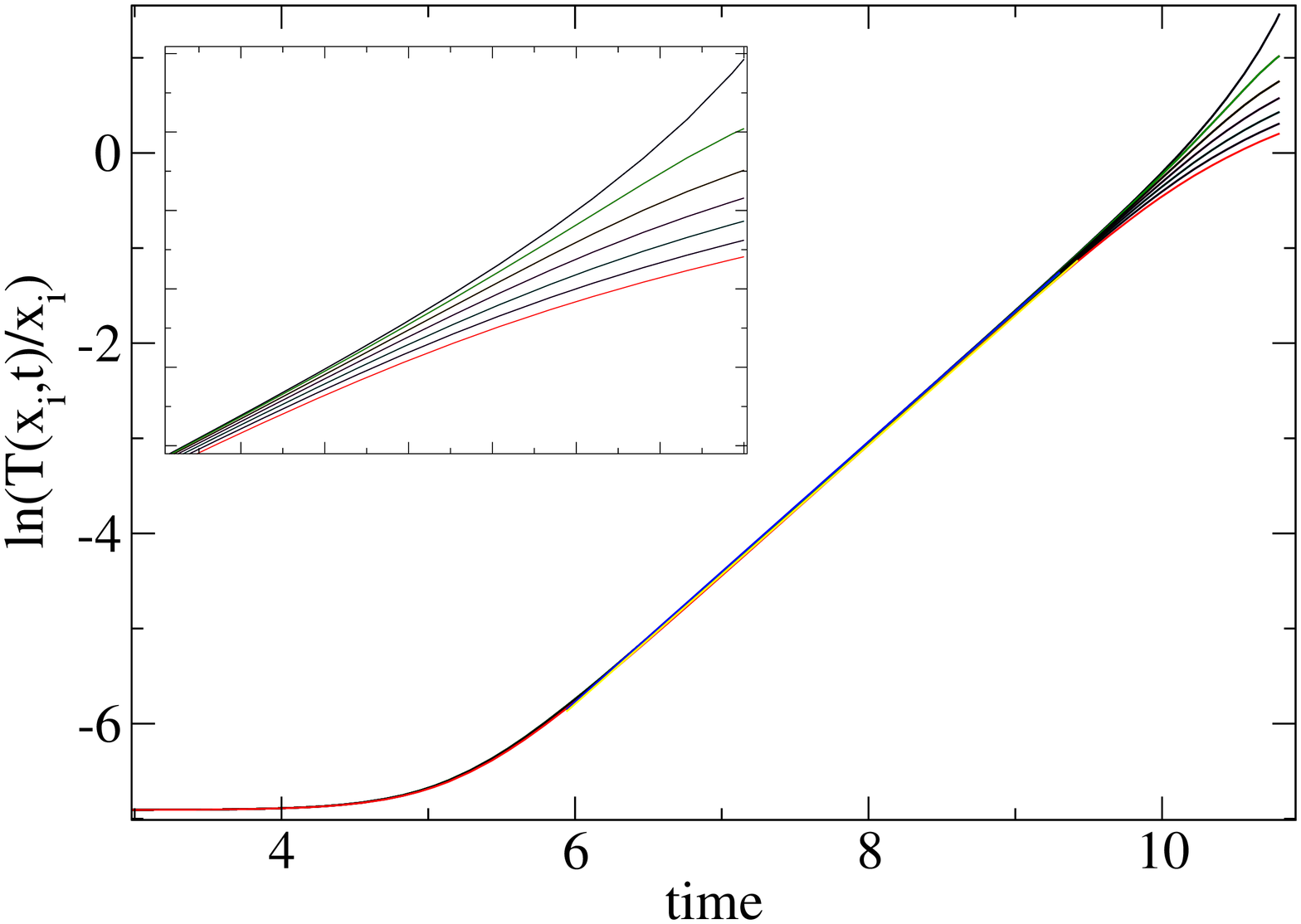}}
{\small
Figure 4. $\ln(T(t,x_i)/x_i)$, which should approximate $q(t)$, 
as a function of time (in units of $a$), for
several lattice sites in the vicinity of a kink.  Inset: close-up of the
large $t$ region.
}
\medskip

{\bf 5.\ Hamilton-Jacobi method.}   The preceding argument is not conclusive
because we have ignored the undetermined function $p(t)$, which could
conceivably soften the behavior of $q(t)$.  Therefore we give an additional
argument which bolsters the conclusion that the slope of the kink diverges in
a finite time.  We use the formalism of \cite{FKS}, in which Hamilton-Jacobi
equations were used to demonstrate the formation of caustics in the tachyon
profile when the initial conditions were inhomogeneous.  At first sight their
formalism appears to be inapplicable to the study of kinks because it is the
leading term in an asymptotic expansion in powers of $e^{-T^2/a^2}$, which is
not small very near the kink. 

However, we have seen that no matter how close a given position $x_i$ may be
to the kink, if we simply wait long enough, the value of $T(t,x_i)$ will
become sufficiently large that we {\it can} expand in $e^{-T^2/a^2}$.  Suppose
we carry out this procedure for two points $x_i$ and $-x_i$ which bracket a
kink at $x=0$; we can then deduce the behavior at the kink by interpolation. 
An interesting aspect of the Hamilton-Jacobi method is that rather than
determining $T(t,x_i)$ for later times, it instead yields $T(t,x_c[t,x_i])$
where $x_c[t,x_i]$ is a characteristic curve that depends on which initial
value $x_i$ was chosen.  We will show that $x_c[t,x_i]$ actually crosses $x=0$
from the right for positive $x_i$, with a nonzero value for $T$ at this
event.  Therefore the tachyon becomes multivalued, since $T=0$ at the origin by
construction.  Its slope must diverge at this moment.

To see this, let us consider some $x_i$ close to the kink, but starting at a
late enough time $t_i\equiv 0$ that $T(t_i,x_i)\equiv T_i \gg a$.  The characteristic
curve associated with this event is 
\beq
	x_c[t,x_i] = x_i - {T_i' t\over \sqrt{1+T_i'^2}} 
\eeq
where $T_i'=T'(t_i,x_i)$, and the solution for $T$ along this curve is
\beq
	T(t,x_c[t,x_i]) = T_i + {t\over \sqrt{1+T_i'^2}} 
\eeq
Clearly, as long as $T'_i\neq 0$ which will be true sufficiently close to the
kink, $x_c$ will cross $x=0$ in a finite time, $t = x_i\sqrt{1+T_i'^2}/T_i'$,
and $T$ will be nonzero.  The procedure is illustrated in Fig.\ 5

\medskip
\centerline{\epsfxsize=4.5in\epsfbox{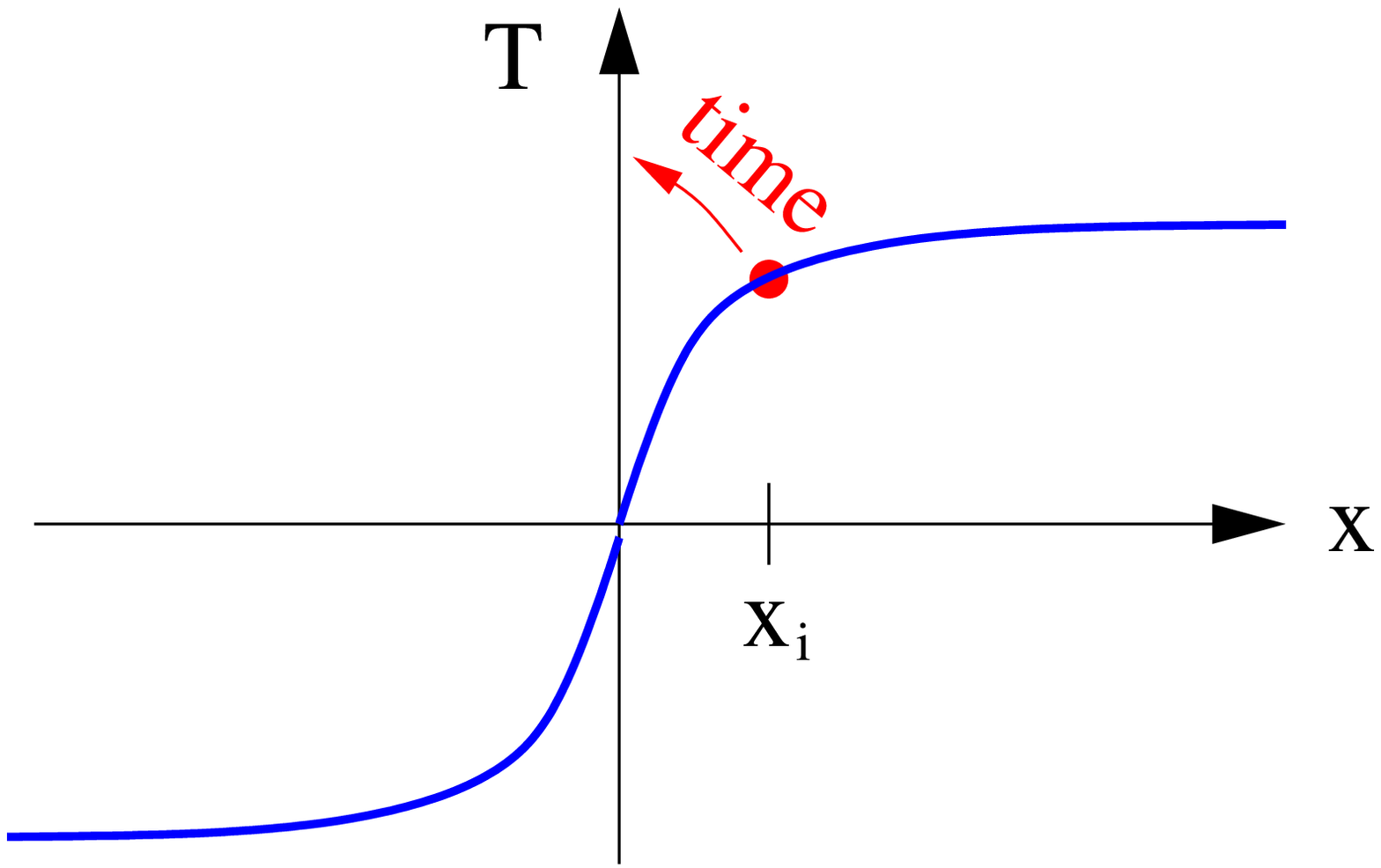}}
{\small
Figure 5. Schematic representation of the temporal evolution of tachyon spatial
profiles in the Hamilton-Jacobi method; the point on the curve at $x_i$ evolves
toward the $T$ axis as shown.}
\medskip

{\bf 6.\ Discussion.}  In this letter we have quantified the time-dependence of tachyon
defect formation, hence formation of a D-brane, in the decay of an unstable nonBPS
brane to a stable one.  Numerical and analytic methods indicate that the profile of
the kink becomes infinitely steep in a finite amount of time of order the string
scale, bringing an end to our ability to follow the evolution.  Notice that $T$ should
continue to roll toward $\infty$ in the bulk at large times.  This situation is 
reminiscent of the problem of caustic formation, where the appearance of cusps also
creates an obstacle to evolving the fields further in time.  

In the case of caustics, the second derivative $T''$ ceases to exist at some point.  It
is tempting to  speculate that this problem is an artifact of the approximations which
led to the action (\ref{action3}).  The BSFT result is only exact for tachyon
configurations with vanishing spatial derivatives higher than first order.  There are
corrections of order $T''^2$ which must exist, but are difficult to compute in BSFT.  The
worldsheet partition function which must be evaluated to find the effective action is no
longer Gaussian when $T''\neq 0$, so such corrections would have to be computed
perturbatively.  So far such calculations have not been done, so we are not able to test
the conjecture that the new terms cure the caustic problem.  If they do, it also seems
likely that they would ameliorate the problems of kink formation.  A fully  string
theoretic calculation might be needed to settle the issue.

For the problem of reheating, the divergence of the slope of the kink is not necessarily
a serious obstacle to doing computations; if most of the production of gauge bosons on
the final brane takes place before the singularity occurs, then it is not so important. 
Intuitively we might guess that the rapid increase of the kink slope will decrease the
efficiency of reheating however.  The reason is that the bulk motion of the tachyon fluid
should not be influenced much by the presence of the defect if the latter is highly
localized.  Therefore we might expect that most of the energy in the tachyon condensate
will be released into gravitons, just as if the final state branes were not present. 
This could be an argument for alternative scenarios of brane-antibrane inflation, such
as branes at angles, where the final state brane has the same dimensionality as the
colliding ones \cite{Sen-discussion}.

\bigskip
We thank D.\ Chung, L.\ Kofman, S.\ Minwalla, A.\ Sen and G.\ Shiu for helpful discussions.

\bigskip
{\bf Note Added:}  After this work was completed, we became aware of \cite{Sen-time}, which found
the formation of a singularity in the energy density of the tachyon configuration at a finite time.
Since this was an exact string theoretic calculation, it means that the effect we found is likely
to be real and not an artifact of truncating the expansion in derivatives of the fields, as suggested above.
It was suggested in \cite{Sen-time} that the origin of the singularity is the formation of a delta function
source of brane tension whose presence cannot easily be inferred from the evolution of the field equations alone.
We thank A.\ Sen for these observations.


\end{document}